# Basics of Low-temperature Refrigeration

*A. Alekseev*[1]
Linde AG, Munich, Germany

**Abstract**
This chapter gives an overview of the principles of low temperature refrigeration and the thermodynamics behind it. Basic cryogenic processes - Joule-Thomoson process, Brayton process as well as Claude process - are described and compared. A typical helium laboratory refrigerator based on Claude process is used as a typical example of a low-temperature refrigeration system. A description of the hardware components for helium liquefaction is an important part of this paper, because the design of the main hardware components (compressors, turbines, heat exchangers, pumps, adsorbers, etc.) provides the input for cost calculation, as well as enables to estimate the reliability of the plant and the maintenance expenses. All these numbers are necessary to calculate the economics of a low temperature application.

## 1  General principles of refrigeration

If your espresso is too hot, you just wait a minute, the ambient air cools the coffee and you can enjoy the drink. If your Coca-Cola is too warm, you put some ice into the cup and your drink suddenly becomes colder. All these actions have some similarity: here, a warm object contacts (directly or indirectly) the colder object. This thermal contact is an essential requirement for the cooling.

What happens at the interface between warm and cold objects from a thermodynamic point of view? The heat from the warm object flows to the cold object: from the hot espresso to the colder ambient air, from the coke to the ice, and so on.

This is our experience: we can cool some object if we have some other material/objects available that are already a bit colder. This could be cold water or snow. If really low temperatures are necessary (–100°C or lower), then we use a special device – the so-called 'refrigerator' (or 'cooler').

### 1.1  What is a refrigerator?

For us, it is important to understand that every refrigerator has an area (or a surface) for which the temperature is lower than that of the ambient. For the cooling of an object, we just need to establish thermal contact between this object and the cold area of the refrigerator, and the heat ($\dot{Q}_0$ in Fig. 1) flows from the object to the cold surface and the object becomes colder.

---
[1] alexander.alexeev@arcor.de

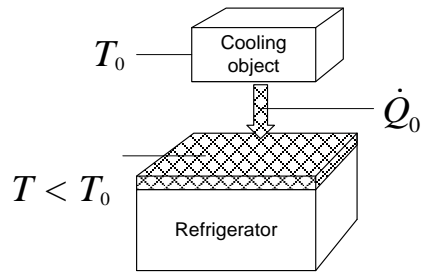

**Fig. 1:** A refrigerator and a cooling object

Let us note three important comments:

**Definition of a cold surface.** Often, this is a real solid cold surface, such as a cold finger in a cryocooler or an evaporator in a household refrigerator. But some refrigerators produce a cold liquid cryogen (liquid helium, liquid neon, or liquid nitrogen), which – while flowing to/through the cooling object (e.g. a superconducting magnet) – evaporates and the vapour is then returned to the refrigerator. In this case, the cryogen liquid surface can be considered as the cold surface.

**Refrigerator/liquefier.** In low-temperature science, we often use another device called a 'liquefier'. A liquefier produces cold liquid that is then drawn off. The thermodynamics is the same for both the refrigerator and the liquefier, but it is useful to start with refrigerator thermodynamics, because it is a bit simpler.

**Cooling capacity.** Some cooling objects do not produce any heat of their own: they just need to be cooled to a defined temperature – and that is all. But the most cooling objects (magnets, sensors, current leads) generate heat continuously during operation, and therefore need permanent cooling. The 'heat' in this case is a permanent heat flow from the cooling object to the refrigerator. This heat flow corresponds to the 'cooling power' or 'cooling capacity' of the given refrigerator.

## 1.2  A refrigerator consumes power

As said above, an important property of every refrigerator is the cold surface for adsorption of the heat from the cooling object.

Another very important property is that it needs driving power – every refrigerator consumes power (electrical or mechanical). Take a look at the conventional home fridge – you will see the power cable for driving the small compressor behind the cabinet.

## 1.3  The first law of thermodynamics for a refrigerator

I expect that you are aware of the first law of thermodynamics – the conservation of energy principle. Inelegantly stated, the first law says that what goes into a system must either come out or accumulate. For steady-state conditions (without accumulation inside the system), the sum of the heat flows going into the system is equal to the sum of the outlet energy streams. An energy stream is, for example, a heat flow or mechanical power for driving a compressor, pump, and so on. All material streams consist of some thermal energy (temperature) or mechanical energy (pressure), and therefore the thermomechanical energies of material streams have to be taken into account by means of enthalpy flow. We will try to apply this law to our refrigerator, working in steady-state conditions:

– two energy streams, the heat flow $\dot{Q}_0$ from the cooling object and the electric power $P$, go into the system;

– 'steady-state' means that there is no any accumulation inside this system.

Consequently, according to the first law, at least one additional 'invisible' energy flow must exist, and this heat flow leaves our refrigerator – it goes out from the system – just to balance the system. This energy flow does indeed exist. It is the waste heat shown in Fig. 2. The usual abbreviation for this heat flow is $\dot{Q}_{amb}$. The first law for a refrigerator can be therefore expressed as

$$\dot{Q}_0 + P = \dot{Q}_{amb}. \tag{1}$$

Every refrigerator produces waste heat, and this heat is rejected to the ambient. At the rear of the household fridge, you can find a black grid, made from up of tubes. This device is warm if the compressor is running. By means of this surface, heat will be removed from the fridge and transferred to the ambient air. Large refrigerators, such as the helium systems at LHC/CERN, use another method – they are cooled by water. In this case, the waste heat is transfered to the cooling water. Small-scale helium refrigerators for laboratories are usually air-cooled.

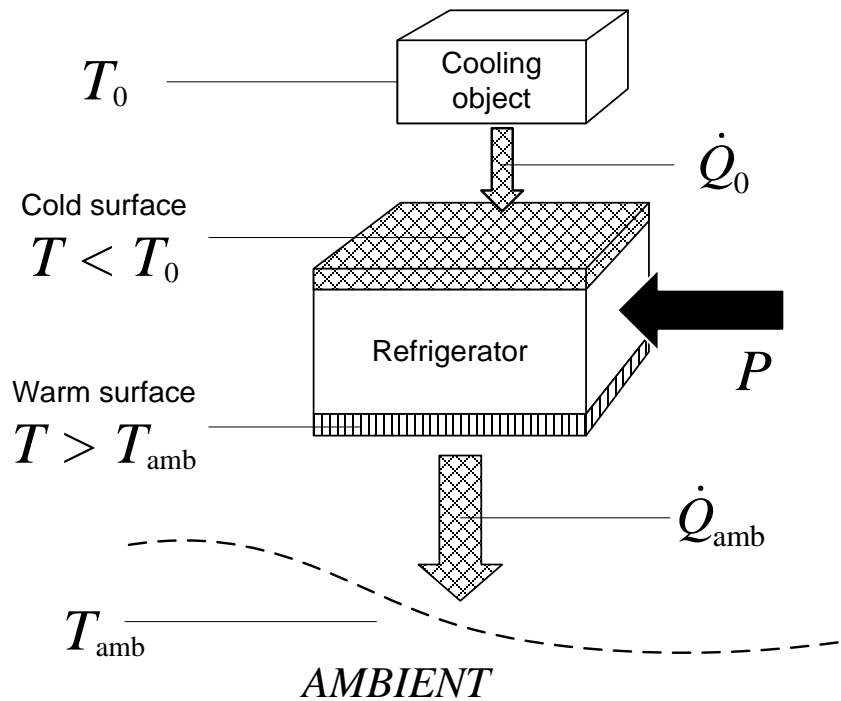

**Fig. 2:** The first law of thermodynamics for a refrigerator

*Summary*

Every refrigerator has at least three interfaces:

- a 'cold surface' to receive and absorb the heat from the cooling object (to cool the cooling object);
- a 'warm surface' to reject the waste heat to the ambient; and
- 'power' fed into the system.

It is important to repeat that the temperature of the 'cold surface' is lower than the temperature of the cooling object, because heat can only flow from a warm object to a cold object.

For the same reason, the temperature of the 'warm surface' is higher than that of the ambient, and the waste heat can flow from this hot surface to the ambient air, which is colder.

## 1.4 The analogy between a refrigerator and a water pump

Hans Quack [1] once found some analogies between a refrigerator and a water pump, pumping the water from a deep-water source (deepness $H_0$ in Fig. 3) to the Earth's surface.

- The pump consumes some power $P$ (it is usually driven by an electric motor).
- It pumps the water from the deep source to the ambient.
- The suction (inlet) nozzle of the water pump is located a little lower in the source than the liquid level – this is necessary to guarantee continuous flow to the pump.
- The pressure at the discharge nozzle is a little higher than ambient pressure – otherwise, water could not flow out.

The refrigerator works in a very similar way.

- The heat $\dot{Q}_0$ flows from the cooling object to the cold surface of the refrigerator.
- The refrigerator lifts (elevates) this heat from the cooling temperature to a temperature a little higher than that of the ambient – this process requires some electrical or mechanical power $P$.
- The heat pumped to the temperature higher than that of the ambient (which becomes $\dot{Q}_{amb}$) is rejected to the ambient.

Some engineers use term 'heat pump' for a refrigerator, because a refrigerator pumps heat from a low temperature to the ambient temperature.

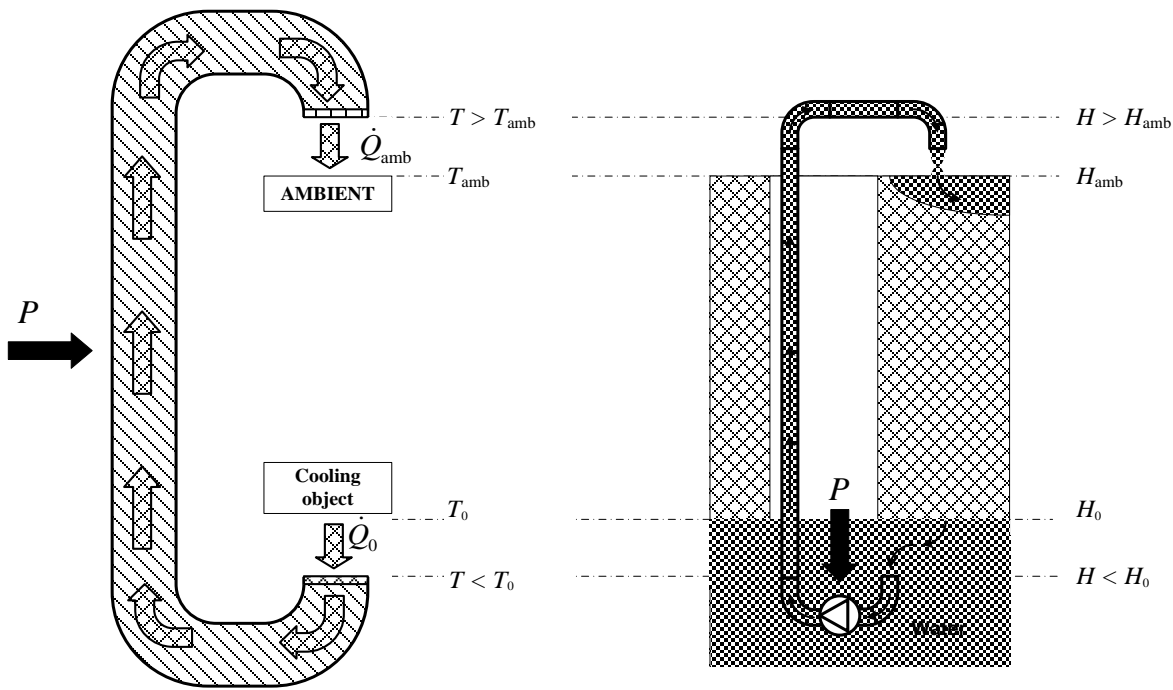

**Fig. 3:** A refrigerator (left) and a water pump (right)

## 1.5 The Carnot equation

The first law equation ($\dot{Q}_0 + P = \dot{Q}_{amb}$) supplies us with some information about the properties of a refrigerator. However, this information is not sufficient for complete analysis of the system. For example, we cannot determine the power consumption of a refrigerator with a cooling capacity of 100 W at 100 K. The first law cannot answer this question. The first law only says that:

- if the power consumption were to be $P = 100$ W, then it would produce waste heat $\dot{Q}_{amb} = 200$ W;

- or if we were to assume a driving power of $P = 1$ W, then the waste heat produced would be $\dot{Q}_{amb} = 101$ W.

But here is the question: is it actually possible to build a refrigerator with a cooling capacity of 100 W at 100 K that consumes 1 W? The first law cannot answer this question. It is actually answered by the second law of thermodynamics, in the form of the so-called Carnot equation, which is valid for an ideal refrigerator – a refrigerator without thermodynamic losses (the Carnot equation in this form was formulated by Rudolf Clausius, but Sadi Carnot developed the thermodynamic model – the so-called 'Carnot engine' – used later by Clausius):

$$P_{MIN} = \dot{Q}_0 \cdot \frac{T_{amb} - T_0}{T_0}. \qquad (2)$$

According to the Carnot equation:

1. the power required to drive a refrigerator depends linearly on the cooling capacity $\dot{Q}_0$ of the refrigerator: the larger the required cooling capacity, the higher the amount of power that is necessary;

2. the power required to drive a refrigerator depends linearly on the ambient temperature $T_{amb}$: the higher the ambient temperature, the higher the amount of power that is necessary;

3. the power required to drive a refrigerator depends on the cooling temperature: the lower the cooling temperature, the greater is the power that is required.

These three statements do not need special explanation. But the following statement is a little different:

4. the power required to drive a refrigerator depends on the cooling temperature: the lower the cooling temperature, the more power is required – and *this function is strongly non-linear; it is 1/x*.

According to the Carnot equation: to produce the 100 W cold capacity at 100 K, while assuming an ambient temperature of 300 K = 27°C, the refrigerator would need at least 200 W:

$$P_{MIN} = 100 \text{ W} \cdot \frac{300 \text{ K} - 100 \text{ K}}{100 \text{ K}} = 200 \text{ W}.$$

This means that a 100 W at 100 K refrigerator with a driving power of less than 200 W cannot exist in nature:

$$P \geq \dot{Q}_0 \frac{T_{amb} - T_0}{T_0}. \qquad (3)$$

## 1.6 A comparison with the water pump and/or vacuum pump

In previous section we discussed the functional analogy between a refrigerator and a water pump. However, *in energy terms, these devices are very different*. This is because the power consumption of the water pump depends on the *pressure difference* between the inlet and the outlet. The power consumption of a refrigerator depends on the *ratio* of the ambient temperature to the cooling temperature $T_{amb}/T_0$. For cooling temperatures below 1 K, the ratio $T_{amb}/T_0$ becomes higher than 300 and the specific power consumption becomes gigantic. Therefore, as you sometimes hear, we are told that absolute zero (0 K) is inaccessible/unreachable. Why? Because it requires a lot of energy.

We use sometimes cooling temperatures in the microkelvin region, but either for very small cooling capacities (usually in the mW range) or for very short periods of time. Otherwise, gigantic amounts of driving power are required.

## 1.7 Minimal power consumption for some typical temperatures

The Carnot equation helps us to estimate the minimal power consumption of refrigerators for typical temperature ranges. Table 1 gives values for the most popular cryogenic temperatures corresponding to liquid helium, liquid hydrogen, and liquid nitrogen, for a cooling capacity of 100 W.

**Table 1:** The minimal power consumption for some typical temperatures/cryogens

|  | Liquid nitrogen (LIN) | Liquid hydrogen (LH2) | Liquid helium (LHe) |
|---|---|---|---|
| Cooling temperature $T_0$, K | 77.4 | 20.1 | 4.2 |
| Required cooling capacity $\dot{Q}_0$, W | 100 | 100 | 100 |
| Minimum power requirement $P$, W | 288 | 1393 | 7043 |
| $P/\dot{Q}_0$ | 2.9 | 13.9 | 70.4 |

The best refrigerator (an ideal refrigerator) producing cold at the liquid nitrogen temperature level (77 K, or about –200°C) consumes at least 288 W. An ideal 100 W refrigerator working at the liquid hydrogen level (20 K, or –250°C) needs approximately 1400 W, a value that is four times higher. And a 100 W refrigerator for liquid helium temperatures (4 K, or –269°C) consumes at least 7 kW.

*Exercise: a helium refrigerator for 1.8 K*

A laboratory has an older helium system with a cooling capacity of 300 W at 4.5 K. The scientists would like to upgrade this system and produce the same capacity, but at 1.8 K. The difference between 4.5 K and 1.8 K is only 2.7 K. This seems to be very small and the expectation is that such a kind of revamp/upgrade would not be very costly, but would have a huge positive impact on the capacity of the laboratory. Just to be on the safe side, we would like to check what this change (from 4.5 K to 1.8 K) means for the power consumption of the refrigerator.

For example, we can determine the theoretical minimum power consumption of the existing system using the Carnot equation and an ambient temperature of 293 K ~ 20 °C:

$$P_{MIN} = 300 \text{ W} \times (293 \text{ K} - 4.5 \text{ K})/4.5 \text{ K} = 19\,233 \text{ W} = 19.2 \text{ kW}.$$

We can then calculate the theoretical minimum power consumption for the future 1.8 K system and compare it with the value for the existing system:

$$P_{MIN} = 300 \text{ W} \times (293 \text{ K} - 1.8 \text{ K})/1.8 \text{ K} = 45\,833 \text{ W} = 45.8 \text{ kW}.$$

Surprisingly, the new value of ~46 kW is essentially higher – by a factor of 2.4. This is a dramatic change. Essentially, it is higher than expected. This revamp from 4.5 to 1.8 K essentially means higher energy consumption and probably a very high cost.

## 1.8 The coefficient of performance, COP

The value of the coefficient of performance (*COP*) can be used to characterize the efficiency of refrigerator systems; it is defined as the ratio of the cold capacity to the driving power (or 'what you get'/'what you pay for'):

$$COP = \dot{Q}_0/P. \qquad (4)$$

If a helium refrigerator produces 300 W and consumes 75 kW, the *COP* amounts to the following:

$$COP = \dot{Q}_0/P = (0.3 \text{ kW}) / (75 \text{ kW}) = 0.004 = 0.4\%.$$

It is really difficult to say that this system is a high-efficiency system, because a *COP* value of 0.4% seems to be very small.

Just to get a feeling for this, we can try to estimate the *COP* for an ideal refrigerator with the same cooling capacity at the same temperature level using the Carnot equation. From the previous exercise (4.5 K refrigerator versus 1.8 K), we know that a 300 W at 4.5 K refrigerator needs at least 19.2 kW. The *COP* for this system amounts to:

$$COP_{MAX} = \dot{Q}_0 / P_{MIN} = 0.3 \text{ kW} / 19.2 \text{ kW} = 0.0156 = 1.6\%.$$

This is amazing: the best helium refrigerator (allowed by nature in any circumstances) has a *COP* of only 1.6%.

From this, we can learn that the *COP* for a liquid helium system is always very small and it is difficult to compare helium systems using the *COP* number only.

## 1.9 Carnot efficiency

Another value used for the characterization of efficiency is the so-called Carnot efficiency – or Carnot Fraction (*CF*) or Figure Of Merit (*FOM*). This is defined as the *COP* of a real refrigerator divided by the *COP* of an ideal refrigerator (calculated by means of the Carnot equation):

$$CF = COP/COP_{CARNOT}. \qquad (5)$$

This value makes more sense, because it gives us some feeling about potential improvements. For our 300 W cooling capacity at 4.5 K:

$$\eta_e = 0.4\% / 1.6\% = 0.25 = 25\%,$$

which is a very high efficiency for a 4.5 K temperature level (see Fig. 4 for comparison with existing cooling systems).

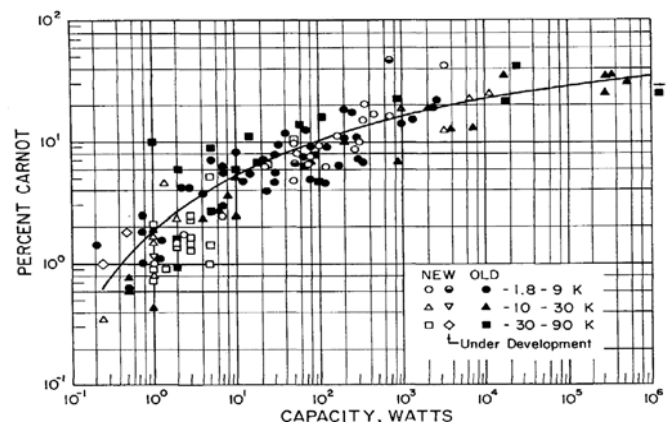

**Fig. 4:** The Carnot efficiency for existing cryogenic systems. Reproduced from [2]

## 2  Cooling effects

Up to now, we have treated the refrigerator itself as a kind of black box. Now, we will try to look inside this black box. What happens inside the refrigerator? How does it work? First, we will look at the available cooling effects. What is necessary to lower the temperature of a gas?

### 2.1  Joule–Thomson expansion

To begin with, let us consider the very simple experiment shown in Fig. 5:

- we take a conventional gas cylinder (50 l) filled with 200 bar nitrogen at room temperature;
- we open the valve carefully; and
- nitrogen flows into the ambient.

The pressure inside the bottle is high (200 bar), while the pressure outside is atmospheric pressure (approximately 1 bar abs). Therefore the gas expands from the high-pressure zone to the low-pressure zone by flowing through the valve. No heat flows, nor any mechanical/electrical energy flows, are used during this experiment. However, if we measure the temperature of the gas at the outlet, we will realize that it is lower than room temperature, the temperature difference being approximately 30°C.

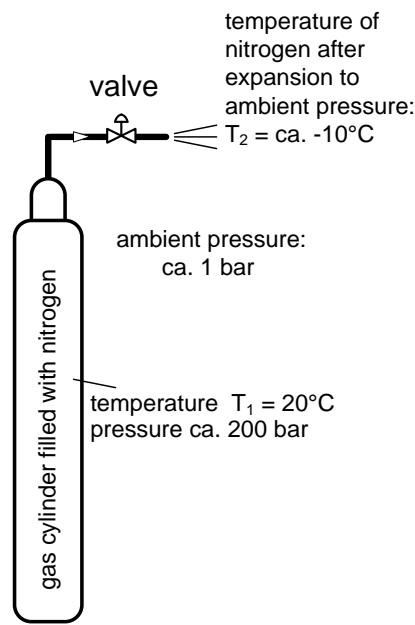

**Fig. 5:** The experimental set-up for the Joule–Thomson cooling demonstration

This temperature change (of a gas when it is forced through a valve or other resistance from higher pressure to lower pressure) $\Delta T = T_1 - T_2$ is the so-called 'Joule-Thomson effect'. This procedure (the expansion of a gas in a throttle valve, orifice, capillary, porous plug, or other pressure resistance) is the called 'throttling' or 'Joule Thomson expansion'. The Joule-Thomson effect is characterized by the so-called 'Joule-Thomson coefficient' $\mu$, which is derived as follows:

$$\mu = \left(\frac{\partial T}{\partial p}\right)_{h=\text{const}}.$$

It is very important for an understanding of Joule-Thomson throttling that there are no changes in the energy of the gas during the throttling procedure – no heat flows, nor any mechanical/electrical energy, go into the system or leave the system during the throttling. Therefore the thermomechanical energy of the gas – the enthalpy of the gas – during the throttling process is constant: the enthalpy at the inlet into the valve and the enthalpy at the outlet are identical.

The Joule-Thomson coefficient is a material property, like the density, the specific heat capacity or any other property. It depends on thermodynamic conditions such as pressure, temperature, and phase state. And, of course, it differs from substance to substance: if we were to repeat our experiment with helium (expanding from high pressure to ambient pressure), the temperature at the outlet would be a little higher than in the bottle. Further, the temperature drop would essentially be smaller compared to nitrogen.

The Joule-Thomson coefficient is a property of real gases, which means gases that cannot be described adequately by the ideal gas equation ($pV = RT$). Theoretically, the enthalpy of an ideal gas (which only exists theoretically) does not depend on the pressure. Therefore the virtual throttling of an ideal gas does not cause any temperature drop. And in reality, for example, nitrogen (or ambient air) at lower pressures (<10 bar) and ambient temperature behaves like an ideal gas, and the throttling of this nitrogen to lower pressure leads to a negligible small temperature change. However, at high pressure (100–200 bar) the properties of nitrogen differ from ideal gas behaviour, and this becomes visible in the temperature change during throttling.

In Fig. 6, the nitrogen expanding process is shown on a temperature–entropy diagram. Point '1' corresponds to the state of the nitrogen in the bottle at the beginning of the experiment: ambient temperature 300 K and pressure 200 bar.

The line, which describes the expansion process in a valve, is the line of constant enthalpy. This is because the enthalpy (heat content) of the gas remains constant: we do not see any heat flows into or out of the expanding gas. At the outlet of the valve, the gas has the same heat content as ahead of the valve, but the pressure corresponds to the ambient pressure. This point is marked by '2'. You can see from Fig. 6 that the temperature difference indeed amounts to ~30 K.

We can pre-cool the bottle with nitrogen to 200 K (–70°C, corresponding to point '3') and repeat the experiment. At the corresponding outlet state, marked here with '4', the temperature difference is essentially higher and amounts to more than 70 K.

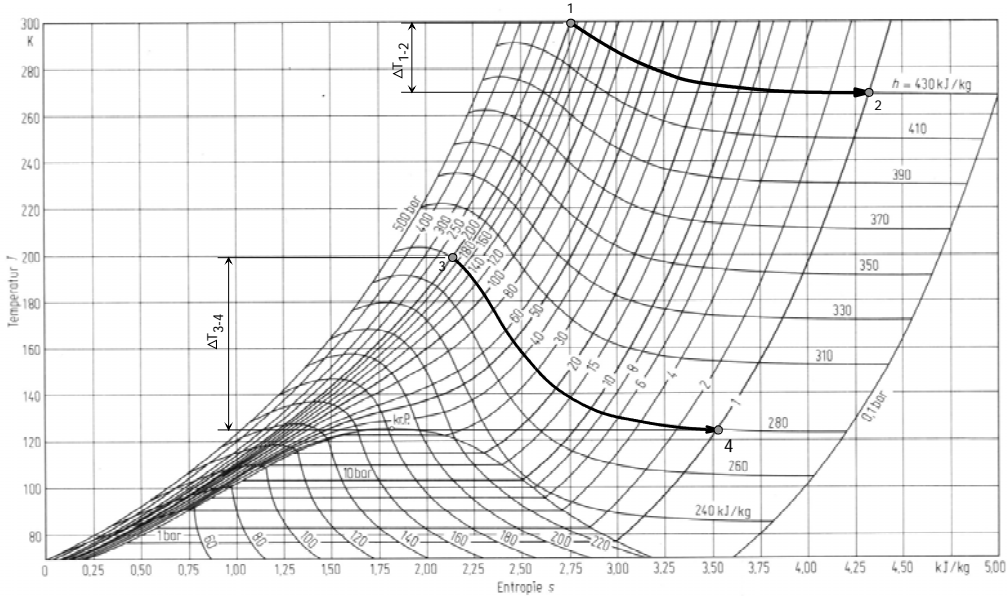

**Fig. 6:** Joule-Thomson expansion of $N_2$ in a temperature–entropy ($T$–$s$) diagram. The $T$–$s$ diagram is reproduced from [6].

The Joule–Thomson coefficient is relatively high at the conditions close to the two-phase area, and it is relatively small at higher temperatures and lower pressures.

The temperature–entropy diagram for helium shown in Fig. 7, for temperatures below 10 K, looks very similar to the diagram for nitrogen. We can see that it is not possible to produce liquid droplets at the outlet by throttling of helium if the temperature before the throttle is higher than 7.5–8.0 K. The liquefaction of helium can only happen below this temperature.

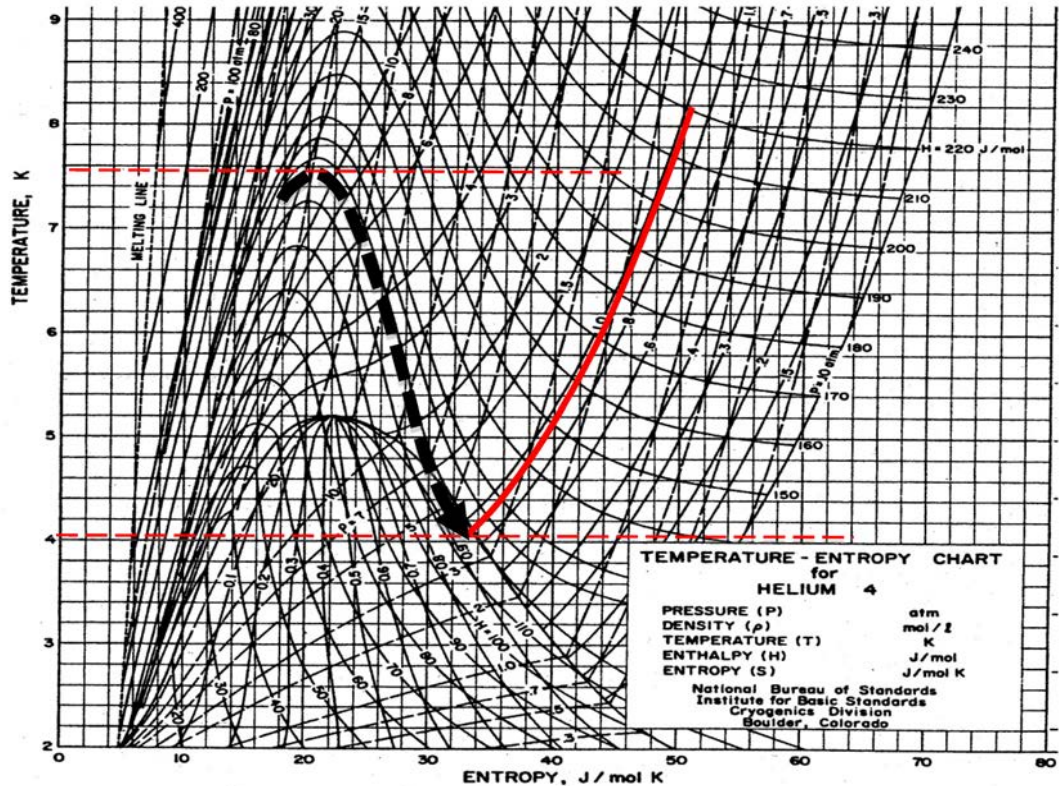

**Fig. 7:** Throttling of helium in a temperature-entropy diagram (dashed line)

## 2.2 Expansion in a turbine

The Joule–Thomson expansion is a very simple method: all you need is a valve or some other pressure resistance such as an orifice or capillary. But the transformation of pressure with temperature change here is inefficient, especially in the gaseous area, because of the small Joule–Thomson coefficient.

From the power generation area, we know another kind of gas expansion – in a turbine. Here, the hot pressurized gas (or steam) expands in a gas turbine (or a steam turbine) to low pressure. The turbine drives an electrical generator (see Fig. 8). During the expansion, the temperature of the gas is reduced. From the power generation point of view, this is only a secondary effect, because the goal of the expansion process (in a power generation plant) is to produce as much power as possible. However, this decrease of temperature effect is important for refrigeration.

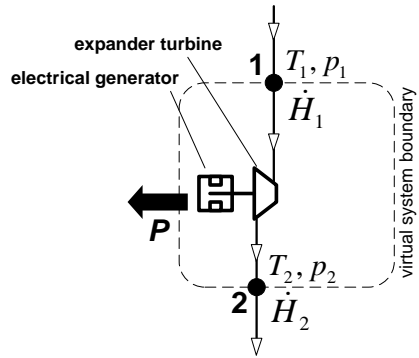

**Fig. 8:** Expansion in a turbine

The important thing is that the gas turbine produces some power: the more power produced, the more efficient is the cooling process. This is because of the first law of thermodynamics: the gas expanding in the turbine does work (produces mechanical power), and its energy content at the outlet (enthalpy flow $\dot{H}_2$) is lower than at the inlet to the turbine (enthalpy flow $\dot{H}_1$) due to the power $P$ produced:

$$\dot{H}_2 = \dot{H}_1 - P,$$
$$\text{if } P > 0, \text{ then } \dot{H}_2 < \dot{H}_1$$

and therefore the temperature at the outlet $T_2$ is lower than the temperature at the inlet to the turbine $T_1$:

$$T_2 < T_1.$$

An ideal expanding process is shown in Fig. 9, in a temperature–entropy ($T$–$s$) diagram. You can see that the expansion of nitrogen from 200 bar and ambient temperature to 10 bar (not 1 bar, but 10 bar) means a temperature change of more than 140 K. If it is pre-cooled to 200 bar, then the temperature change is about 80 K. This is indeed much more efficient than the Joule-Thomson expansion.

The expansion of gas in a turbine works better in the gaseous area, because the heat capacity in the gaseous area is higher.

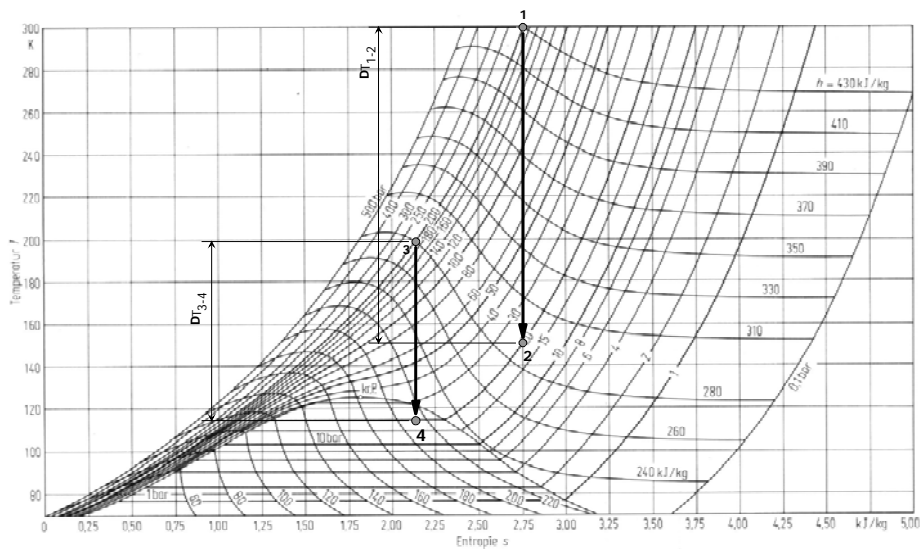

**Fig. 9:** Expansion in an ideal gas turbine expander, shown in a $T$–$s$ diagram. The $T$–$s$ diagram is reproduced from [6].

## 3 Basic cycles

### 3.1 The Joule–Thomson refrigerator

#### 3.1.1 Description of the process

The Joule-Thomson process (the simple Linde process) was developed by Carl Linde for the liquefaction of air in 1895, and can be used for refrigeration at the nitrogen temperature level. It is a really simple process and it always works.

The main hardware components required for realization of this process are a multistage compressor with intercoolers and an aftercooler, a counter-current heat exchanger (Joule-Thomson heat exchanger) and a throttling valve (Joule-Thomson valve), as shown in Fig. 10.

The most important part is the heat exchanger. This piece of equipment will be discussed later in Section 4.2.3. The working principle can be explained with the help of Figs. 23 and 24, which show the so-called 'plate-fin heat exchanger'. This device consists of several channels divided by aluminium sheets, inside of the channels you can see corrugated fins. The warm fluid flows through the small channels, while the cold fluid flows through the other channel in the counter-current direction. This cools the heat exchanger surface and therefore the warm gas on other side of the surface.

The other important device is a compressor. This machine compresses the gas from 1 bar to ~200 bar. Because it is difficult to compress a gas in one step, it is a multistage machine (usually more than four stages).

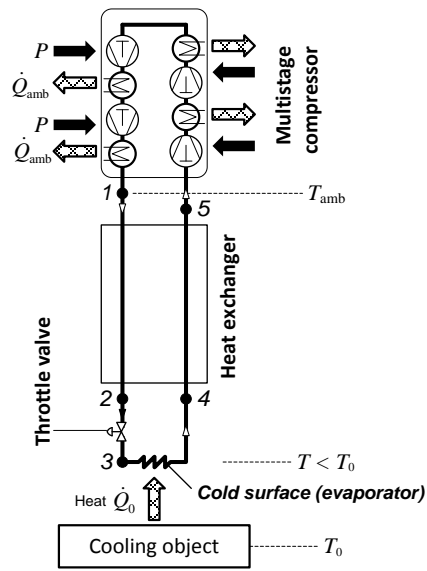

**Fig. 10:** The Joule-Thomson process

#### 3.1.2 The cooling-down procedure

What happens in our refrigerator when the compressor starts to work? It compresses the nitrogen to 200 bar, and in the aftercooler this nitrogen will be cooler to temperature close to the ambient temperature by means of cooling air or cooling water. Then, the pressurized nitrogen flows into a heat exchanger.

The first portion of the nitrogen flows through the heat exchanger without any temperature change, because at the beginning the heat exchanger is warm, its temperature being the ambient temperature. Therefore the nitrogen at the outlet of the heat exchanger has the same temperature as at the inlet, while the pressure is a little lower than 200 bar because of some pressure losses in the heat exchanger (point '1' in Fig. 11).

Next, the nitrogen is throttled in the Joule-Thomson valve, the outlet pressure being close to 1 bar (point '3a' in Fig. 11). This process is identical to our experiment with a nitrogen bottle: during this expansion, the temperature dropped by ~30 K.

This first portion of the nitrogen returns to the compressor through the evaporator and heat exchanger. This gas is colder than the gas on the high-pressure side of the heat exchanger, and therefore it cools the warm gas. Consequently, the second portion of the gas arriving at the Joule-Thomson valve has a temperature that is lower than the ambient temperature (point '2b' in Fig. 11).

After this nitrogen has been throttled in the valve to low pressure, the temperature at the outlet becomes a little lower than the temperature of the first portion of the nitrogen (point '3b' in Fig. 11) after throttling. This cold gas flows back through the heat exchanger and cools the warm high-pressure gas further.

In this way, the heat exchanger is cooled further and further, and at the end of the cooling-down process the temperature ahead of the throttle valve is ~160 K (–110°C). If throttling is carried out starting from this temperature, then the nitrogen at the outlet of the valve is so cold that it is partially liquefied. The surface beyond the throttle valve is cold and we can use it to cool some objects or for other applications. The heat from the cooling objects evaporates the nitrogen, and this vapour returns to the compressor through the heat exchanger.

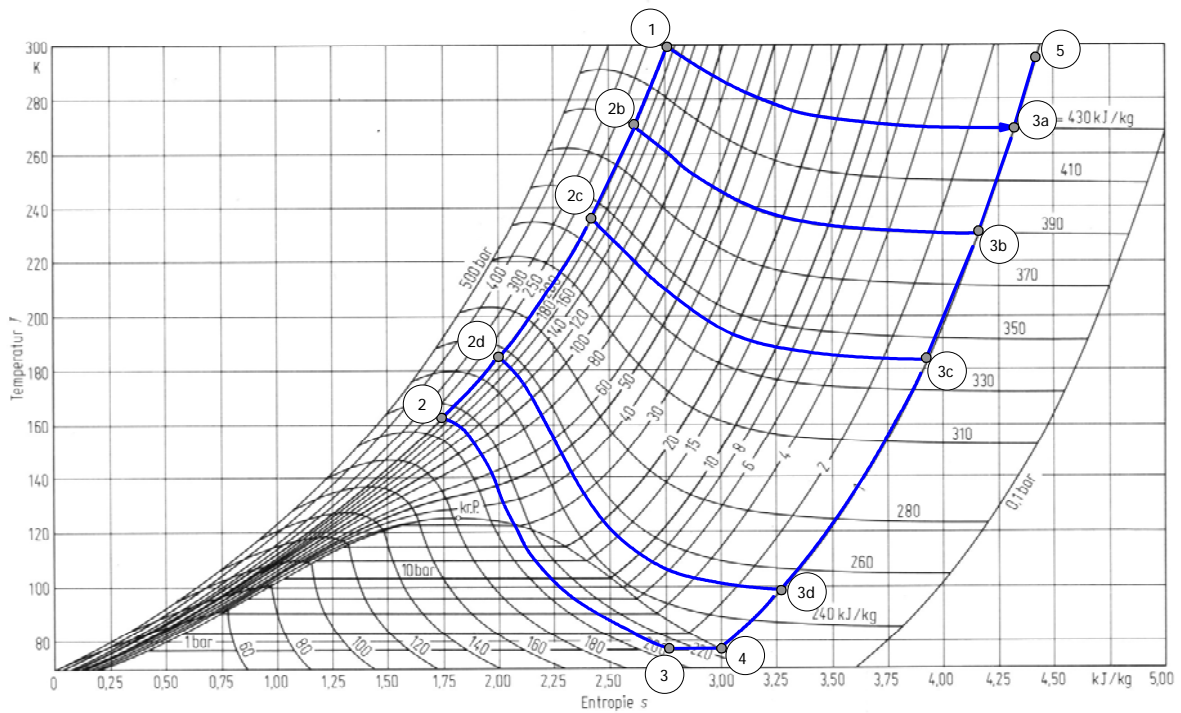

**Fig. 11:** The cooling-down process in a *T–s* diagram. The *T–s* diagram is reproduced from [6]

### *3.1.3 Liquefier versus refrigerator*

The kind of cooling system discussed before is the so-called 'refrigerator'. We can install a small vessel (separator) after the throttle valve as shown in Fig. 12 on the right, and if some nitrogen liquefies after throttling, the liquid will accumulate in this vessel. Usually, 5–7% of the nitrogen flow is liquefied, and the rest (~95%), which is still gaseous, goes through the heat exchanger and cools the high-pressure stream. If some liquid accumulates in this system, we have to feed the same amount of nitrogen into the system just to balance the cycle. The liquid nitrogen can be taken from the system and used. This kind of cycle is known as a 'liquefier'.

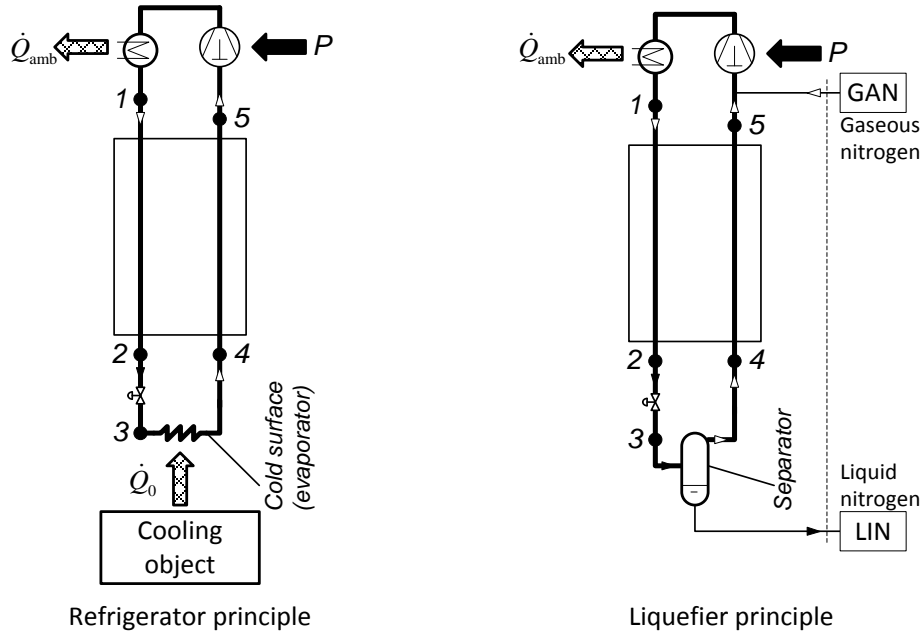

**Fig. 12:** The Joule-Thomson process as a refrigerator or liquefier

### 3.1.4   *The advantages and disadvantages of the Joule-Thomson process*

The big advantage of a Joule-Thomson refrigerator is its simplicity. It consists of a relatively low number of hardware components, with no moving parts in the coldbox. If the heat exchanger is sized more or less appropriately, it always works.

Another important feature of the Joule-Thomson system is that it produces liquid (liquid nitrogen in case of nitrogen Joule Thomson process). The cryogenic liquids have several advantages as coolants.

1. Nitrogen evaporates at a constant temperature. Therefore it is possible to keep the temperature of cooling objects stable.
2. The density of the liquid is much higher than that of the gas; one can use small cooling channels.
3. The volumetric heat capacity of the liquid is much higher.
4. The heat transfer coefficients for boiling liquid are high.
5. It is possible to collect the liquid for back-up purposes (safety and availability).

The main disadvantage is the relatively high pressure, which means that a high-pressure multistage compressor is required. This should be an oil-free compressor, because some oil particles and/or contaminants, if at low temperature, can freeze out and clog the heat exchanger channels or the Joule-Thomson valve.

Such a high-pressure oil-free compressor is relatively expensive and requires some maintenance every 2000–5000 h. Such compressors are only available for relatively small capacities, so that refrigerators with a capacity above 1 kW need several compressor units and the system then becomes really expensive.

The biggest disadvantage is that the cooling temperature of a Joule-Thomson refrigerator is limited to ~54 K (using a mixture of oxygen and nitrogen). To achieve a lower temperature, cryogens such as helium, neon, and hydrogen, and isotopes of these gases are necessary. At ambient temperature, these gases behave like an ideal gas and therefore the Joule-Thomson coefficient is close to zero, which means that it is impossible to build a Joule-Thomson refrigerator system based only on these gases.

Therefore, the market share of pure Joule-Thomson systems is relatively small.

The relatively new development in Joule-Thomson refrigeration is the so-called 'mixed gas refrigerator'. This looks like the classic nitrogen Joule-Thomson refrigerator. However, it uses a special mixture of gases based on nitrogen, methane, ethylene, propane, butane, and some helium/or neon.

Because of the special properties of this mixture, it is possible to reduce the high pressure in the cycle to below 20 bars. Another feature is that some components of this mixture – for example, propane – can dissolve small amounts of compressor oil contaminants (such as solvents). Therefore it is possible to use small oil-lubricated hermetic compressors from air conditioning systems for this kind of refrigerator. These compressors are cheap and reliable (20 000 h or more); therefore, the whole mixed-gas Joule-Thomson system is not as expensive and has a higher availability.

Units for relatively small cooling capacities of approximately 1 W at 80 K (Cryotiger) and for the biggest cryogenic systems for liquefaction of natural gas (LNG) are commercially available.

### 3.2 The Brayton process

The Brayton cycle is the second basic cryogenic cycle. It looks like the Joule-Thomson cycle, but it uses a turbine instead of the Joule-Thomson valve for gas expansion like shown in Fig. 13.

The Brayton cycle does not need very high pressures, because the expansion in the turbine is much more efficient (in comparison to the Joule-Thomson system). It is possible to work with pressures of 10–12 bar at the compressor outlet and achieve high efficiency. This is the big advantage.

Due to the fact that this process uses expansion of gas in a turbine and this kind of expansion is combined with temperature reduction that is always independent of the cryogen used in the cycle, it is possible to use real low-temperature refrigerants such as helium, neon, and hydrogen and achieve temperatures below the liquid nitrogen level.

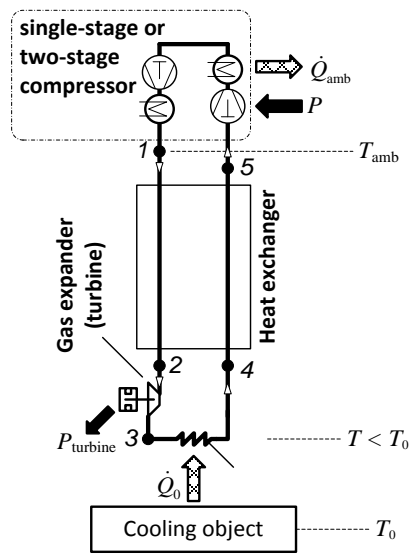

**Fig. 13:** The Brayton process

However, this system also suffers from some disadvantages. The first is that the turbine usually cannot work in the two-phase region. If a partial liquefaction takes place and some of the liquid particles in the turbine fly with a velocity close to the sonic velocity (200–300 m/s), they can damage the turbine wheel. Therefore, the normal Brayton process design is such that any liquid at the turbine outlet is avoided. The consequence is that the classic Brayton cycle cannot produce any liquid: the cold surface of the refrigerator is cooled by cold gas only. Therefore it is difficult to keep the temperature stable.

The second problem is that a turbine, even the smallest with a wheel diameter of 2–3 cm, requires relative large gas flows, and therefore the lower limit for the cooling capacity of a classic Brayton system amounts to ~500 W. It is really difficult to build a turbine with a smaller cooling capacity.

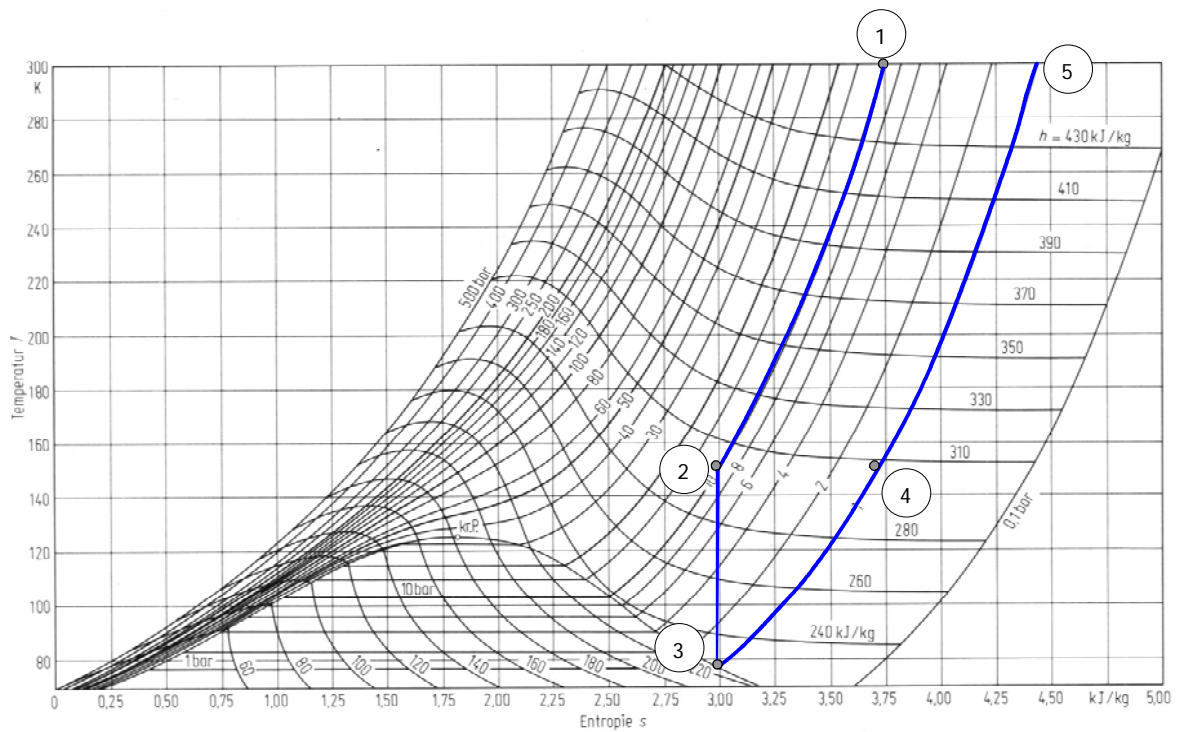

**Fig. 14:** Nitrogen Brayton process in a *T–s* diagram. The *T–s* diagram is reproduced from [6]

### 3.3 The Claude process

You can see that the Joule-Thomson expansion is good for the production of liquids, but it is not really efficient. Expansion in a turbine is very efficient in the gaseous area, but is problematic with regard to the production of liquids. The next cycle – the so-called 'Claude cycle' – is a combination of both the Joule-Thomson and the Brayton processes, and is shown in Fig. 15. The low-temperature part (in grey) looks like the Joule-Thomson cycle and is the Joule-Thomson stage of the Claude cycle.

The upper part (1a → 2a → 3a → 4a → 5a) is similar to the Brayton cycle. Both cycles are driven by the same single compressor. The Brayton cycle used here as a kind of pre-cooling cycle: it is required to reduce the temperature at the inlet into the Joule-Thomson stage. The compressor pressure is usually a little lower than in a classic Joule-Thomson cycle, but higher than in the classic Brayton cycle.

The Claude process combines the advantages of the Joule-Thomson and Brayton cycles. It can produce liquid because of the Joule-Thomson stage and it is very efficient because of super-efficient pre-cooling based on expansion in a turbine.

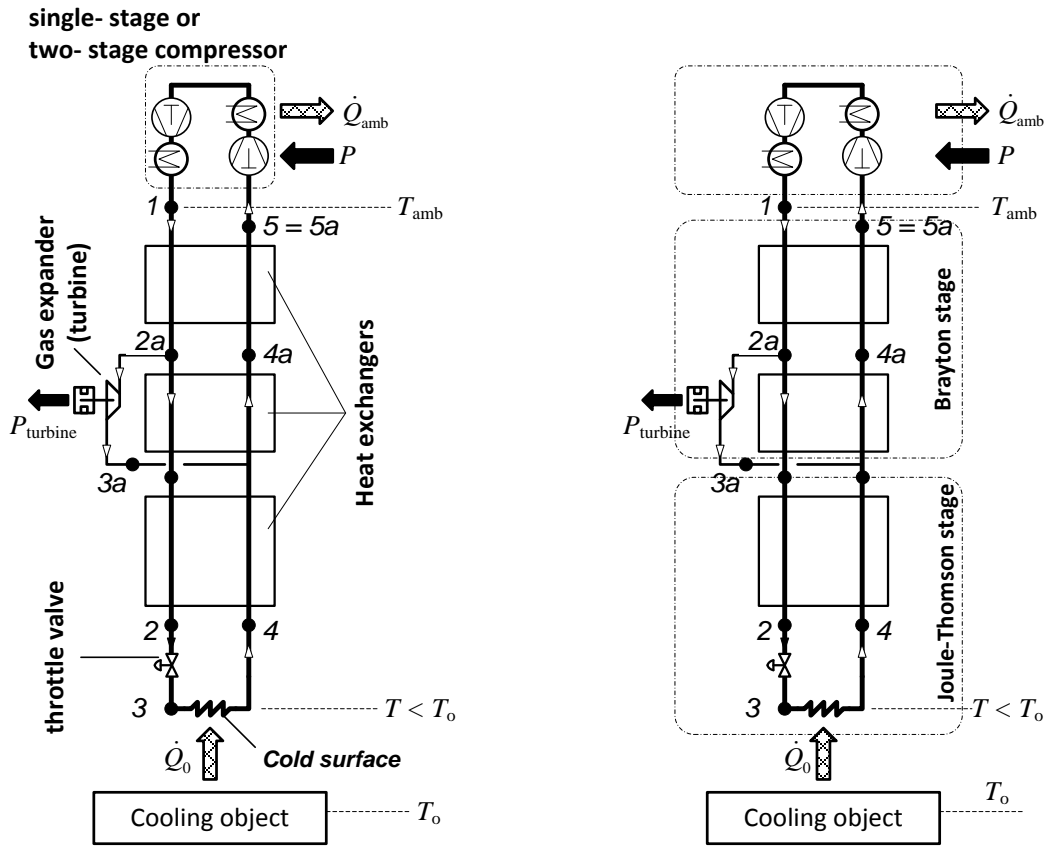

**Fig. 15:** The Claude process

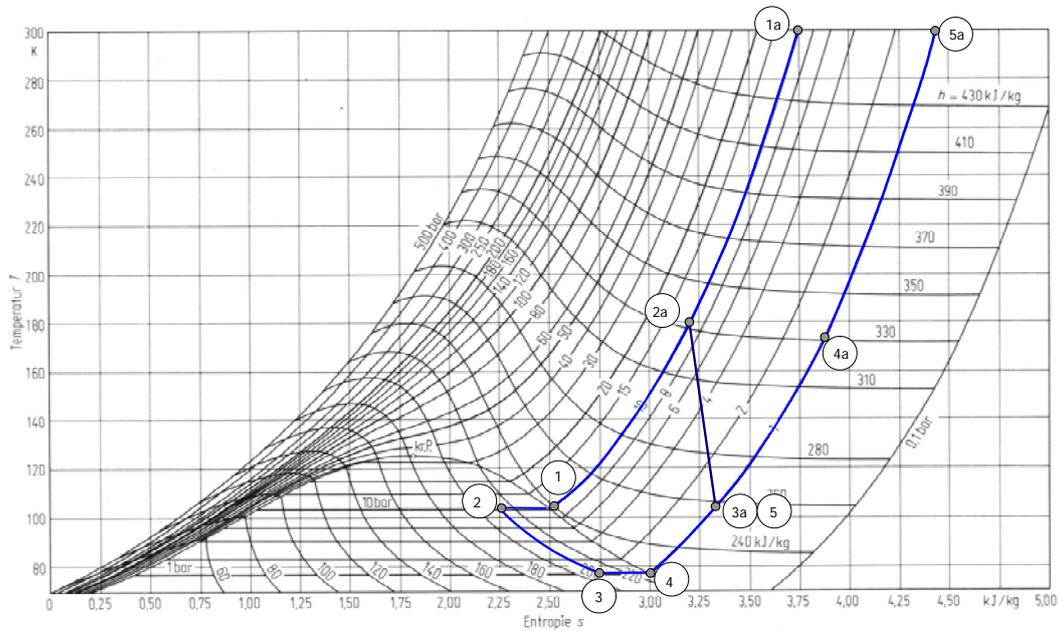

**Fig. 16:** The Claude process in a *T*–*s* diagram. The *T*–*s* diagram is reproduced from [6]

# 4 Helium refrigeration

Table 2 summarizes typical applications for helium refrigeration and provides typical values for the corresponding cooling capacity and cooling temperature.

- Bulk helium liquefiers are used for the separation of helium gas from helium-containing natural gas and for its further liquefaction (for transportation purposes) on-site in the field. However, the current market size is quite small and it is limited to less than three systems every 10 years.
- The market size for a compact laboratory helium liquefier is larger. This kind of system is very cost-efficient because of the high level of standardization and the use of low-cost oil-lubricated screw compressors. This system has become more of a commodity – every other university in Europe has its own small helium liquefier.
- Helium refrigerators for high-energy physics (for the cooling of magnets, cavities, and other objects) are most complex helium systems, producing a huge cooling capacity (up to 25 kW) at different temperature levels down to the extremely low temperature of 1.8 K. Considering their high power consumption, they are designed to achieve the highest efficiency in order to reduce the customer's operating cost.

**Table 2:** Typical applications for helium cryogenics (Claude process)

| Application | Cooling temperature | Cryogen | Cooling capacity |
|---|---|---|---|
| Bulk helium liquefier | 4.3 K | He | 1000–4000 l/h |
| Laboratory helium liquefier | 4.2–4.6 K | He | 20–200 l/h |
| High-energy physics | 1.8 K<br>4.4 K<br>80 K | He | 1–20 kW |

## 4.1 The laboratory helium liquefier

### 4.1.1 Process design

Figure 17 shows a typical process flow diagram of a laboratory helium liquefier.

*Compression*

The helium gas is first compressed in a single-stage oil-lubricated compressor to 10–14 bar and then cooled in an aftercooler. The compressor oil is separated by means of of a relatively complex (three- to four-stage) Oil Removal System (ORS).

*Liquid nitrogen pre-cooling*

This compressed gas is cooled in the E3100A heat exchanger to ~80 K. This cooling can be supported by external cooling sources such as liquid nitrogen, if available. As the liquid nitrogen flows through the separate channels of the heat exchanger, it evaporates and cools the warm helium gas.

From the process design point of view, the liquid nitrogen pre-cooling can be replaced by an additional Brayton stage (with expansion turbine[s]) at this temperature level. However, pre-cooling with liquid nitrogen is widely used for the following reasons:

(1) low capital expenditure (facilitated only by additional channels in the heat exchanger, along with liquid nitrogen supply hardware, instead of using highly sophisticated and therefore expensive cryogenic turbine expanders); and

(2) better overall liquefaction efficiency, since liquid nitrogen is produced by an air separation plant in a more efficient way compared to cold production in helium plants – this is because large-scale turbocompressors and expansion turbines used in air separation plants have a higher efficiency.[2]

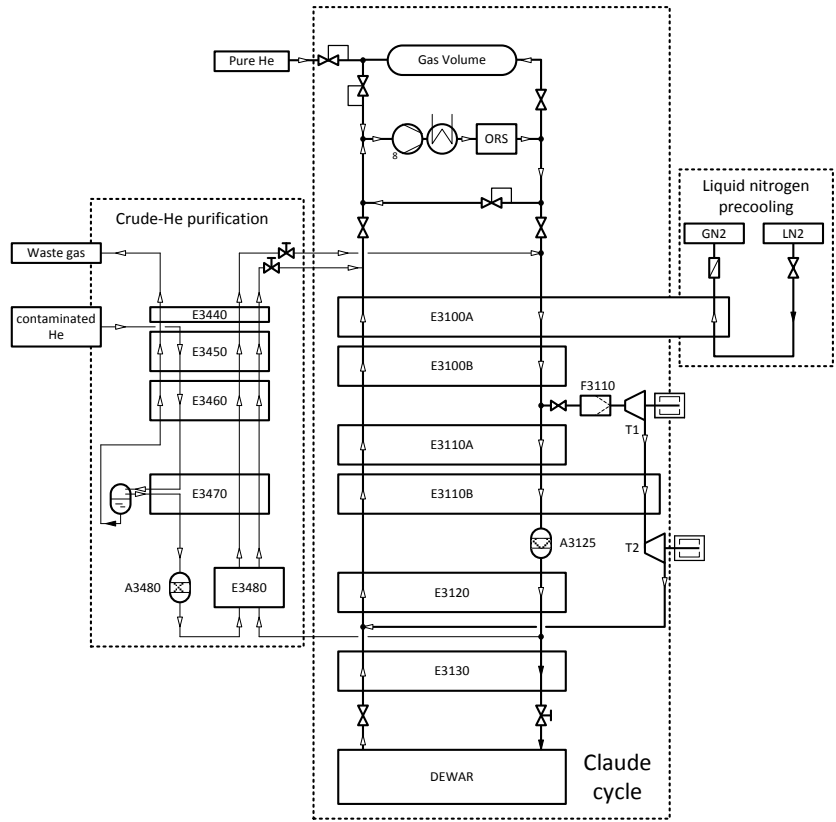

**Fig. 17:** The Process Flow Diagram (PFD) of a laboratory helium liquefier

*The Claude part*

- The combination of the E3100B, E3110A/B, E3120, and E3130 heat exchangers and the T1/T2 expansion turbines corresponds to the classical Claude process, described in the previous section.

- The helium gas is cooled in the E3100B heat exchanger and divided into two streams – the so-called 'turbine stream' and the Joule-Thomson stream.

- The turbine stream is cooled by expansion in two turbines to a low pressure (slightly above ambient pressure). It flows through the heat exchangers back to the surge line of the compressor and becomes warm due to heat transfer from the warm Joule-Thomson stream.

- The Joule-Thomson stream is cooled to below 8 K by thermal contact with the cold turbine stream, before being expanded in a Joule-Thomson valve into a two-phase area and fed into the liquid helium storage vessel – the so-called Dewar.

---

[2] The isothermal efficiency of a large-scale turbocompressor is usually about 75%, while the isothermal efficiency of a helium compressor is usually lower than 60%. The isentropic efficiency of an air expansion turbine is typically in the range of 86–91%, while the isentropic efficiency of a helium turbine is usually lower than 82%.

- The vapour fraction, separated from the liquid in the Dewar, flows through all of the heat exchangers, back to the suction line of the compressors; in this way, it cools the warm high-pressure stream and it becomes warm before re-entering the compressors.

*Purification of crude helium*

The helium gas feed is often contaminated with water and air gases such as oxygen and nitrogen (and therefore is often called 'crude gas'). This crude helium is usually purified prior to liquefaction to avoid solidification, which can lead to blockage of heat exchangers and clogging of valves.

This task is fulfilled by a purification unit shown on the left-hand side of Fig. 17. It consists of a dryer (not shown in the figure) and of a low-temperature purifier for the removal of nitrogen and oxygen. The contaminated helium is cooled in the E3450 and E3460 heat exchangers to ~65–70 K; the liquefied portion of the contaminants is separated in a liquid separator (the separated liquid flows through the heat exchangers into the ambient), while the gaseous contaminants are removed by means of the A3480 cold adsorber – a vessel filled with materials that adsorb the nitrogen and oxygen.

The cooling duty required for cooling the feed gas is delivered by a small cold slip stream from the main liquefier cycle; this slip stream cools the crude feed gas stream while flowing through the E3470, E3460, and E3450 heat exchangers, before it re-enters the helium compressor.

The cold adsorber, as well as the dryer, needs regeneration after a certain working period to guarantee the adsorption capability. Therefore, these devices are periodically heated for this purpose (once a week, depending on the feed amount). The collected contaminants are released into the atmosphere.

*Additional hardware components*

Two additional hardware components are shown in Fig. 17:

- filter F3110, for the protection of expansion turbines from solid particles;
- the additional cold adsorber A3125 to protect the lowest part of the cycle from contamination (just for safety, or for long periods of operation).

### 4.1.2   Capacity

The liquefaction capacity is usually visible as a rising liquid level in the Dewar vessel. The condensed liquid displaces the vapour helium from the upper part of the vessel into the main helium cycle. Therefore the liquefaction cycle is actually fed from the two sources simultaneously: from the cold gas derived from the Dewar vessel (typically 10–15% of the liquefaction capacity, depending on the vessel) and from the real external feed.

Here are some typical specific consumption values (per litre of liquefied helium) taken from Ref. [3]:

- 0.22 kW is the theoretical thermodynamic minimal liquefaction energy demand;
- 1.8 kW is the typical energy requirements for a small laboratory liquefier, without liquid nitrogen pre-cooling;
- 0.9 kW is the typical power demand for a small laboratory liquefier equipped with liquid nitrogen pre-cooling, which then requires close to 0.5 l of liquid nitrogen supply;
- optimized large-scale helium plants can achieve the specific value of 0.5 kW.

## 4.2   Key hardware components

The process design is a scientific art on the border between thermodynamics and engineering: it is about how to design the most energy-efficient and cost-efficient system based on the available hardware components: on one hand, the availability of hardware components limits the creativity of the engineers; but on the other hand, the thermodynamic simulations performed for conceptually new processes set the new targets for developing new hardware components or for improving existing hardware.

Several key hardware components have been developed over recent decades to fulfil some of the special requirements of helium refrigeration, such as very the low operational temperature and the related special properties of cryogenic fluids. The main components are compressors, expanders, and heat exchangers.

### 4.2.1 Compression

The compression of helium gas is a very interesting process, because of some of the special properties of this gas:

- helium is a very light substance (molecular weight 4) and has very low density;
- it is a single-atom molecule, which leads to a relatively high isentropic exponent[3] of $\kappa = 1.7$.

#### 4.2.1.1 The impact of the isentropic exponent

The following equations describe an ideal isentropic compression process. The outlet temperature $T_{out}$ depends on the inlet temperature $T_{in}$, the pressure ratio ($p_{out}/p_{in}$) and the isentropic exponent $\kappa$:

$$T_{out} = T_{in} \cdot \left(\frac{p_{out}}{p_{in}}\right)^{\frac{\kappa-1}{\kappa}}, \quad \text{Power} = \dot{M} \cdot R \cdot T_{in} \cdot \frac{\kappa}{\kappa-1}\left[\left(\frac{p_{out}}{p_{in}}\right)^{\frac{\kappa-1}{\kappa}} - 1\right].$$

The impact of the isentropic exponent is illustrated in Table 3. An ideal isentropic single-stage compression from ambient pressure (1.013 bar) to 10 bar is calculated here for a gas flow of $\dot{M} = 100$ mol/s: the inlet temperature is given by $T_{in} = 300$ K. The calculation is done for two different substances – air and helium; that is, for two fluids with differing values of the isentropic exponent ($\kappa = 1.4$ for air).

**Table 3:** The ideal isentropic compression of helium in comparison to the compression of air

|  | Air | Helium |  |
|---|---|---|---|
| $\dot{M}$ | 100 | 100 | mol /s |
| $R$ | 8.31 | 8.31 | J/mol/s |
| $T_{in}$ | 300 | 300 | K |
| *kappa* | 1.4 | 1.7 |  |
| $p_{in}$ | 1.013 | 1.013 | bar |
| $p_{out}$ | 10 | 10 | bar |
| $T_{out}$ | 577 | 770 | K |
| $T_{out}$ | 304 | 497 | °C |
| *Power* | 806 | 949 | kW |

This difference means that the outlet temperature of 497°C is extremely high in the case of helium – approximately 200 degrees higher compared to the case for air ($T_{out}$ = 304°C). The required power of ~950 kW for helium is approximately 20% higher than the compression power for the air (~800 kW). This very simple calculation indicates very high compression temperatures and makes it obvious that an efficient cooling procedure is required during compression of the helium.

---

[3] The isentropic exponent $\kappa$ describes the relation between the pressure and temperature of a gas during an isentropic process. A relatively high value of 1.7 means that a small pressure change (pressure ratio) leads to a large temperature change. For an ideal gas, the isentropic exponent is equal to the 'heat capacity ratio', $\kappa = \gamma = c_p/c_v$; that is, the ratio of the heat capacity at constant pressure, $c_p$, to the heat capacity at constant volume, $c_v$.

*Low density*

The low density of helium is the main reason and explanation for the fact that highly efficient turbocompressors are not applied for helium compression, although this kind of equipment is always used in chemical engineering for the compression of gaseous fluids.

A turbocompressor consists of several compressor stages, and every compressor stage includes three elements: a guide vane, an impeller (wheel), and a diffuser (volute chamber). The gas is accelerated first in the guide vane, but mainly in the impeller (here, the mechanical energy of the gas is transformed into kinetic energy, with an end velocity close to 300 m/s), and after the acceleration the gas is decelerated in the diffuser (a flow channel with a widened cross-section). The kinetic energy of the gas is transferred into potential energy (pressure). The simplified Bernoulli equation,

$$p_1 + \tfrac{1}{2}\rho_1 v_1^2 = p_2 + \tfrac{1}{2}\rho_2 v_2^2,$$

describes this process, where $p_1$, $p_2$, $\rho_1$, $\rho_2$, $v_1$, and $v_2$ are pressures, densities, and velocities at the inlet and outlet of the diffuser.

The pressure difference achieved in a single compression stage therefore depends on the velocity of the impeller as well as on the gas density. This means that a relatively high pressure difference (and therefore pressure ratio) can be achieved by compressing a heavy gas (such as argon, krypton, or xenon). However, the pressure ratio for light gases such as helium is limited to 1.05–1.2. This means that several compression stages are necessary, instead of a single stage as for heavy gases. Taking into account a corresponding number of intermediate coolers (necessary after every stage because of the high compression temperature) along with pressure losses and so on, a turbomachine for helium would become very complex and expensive.

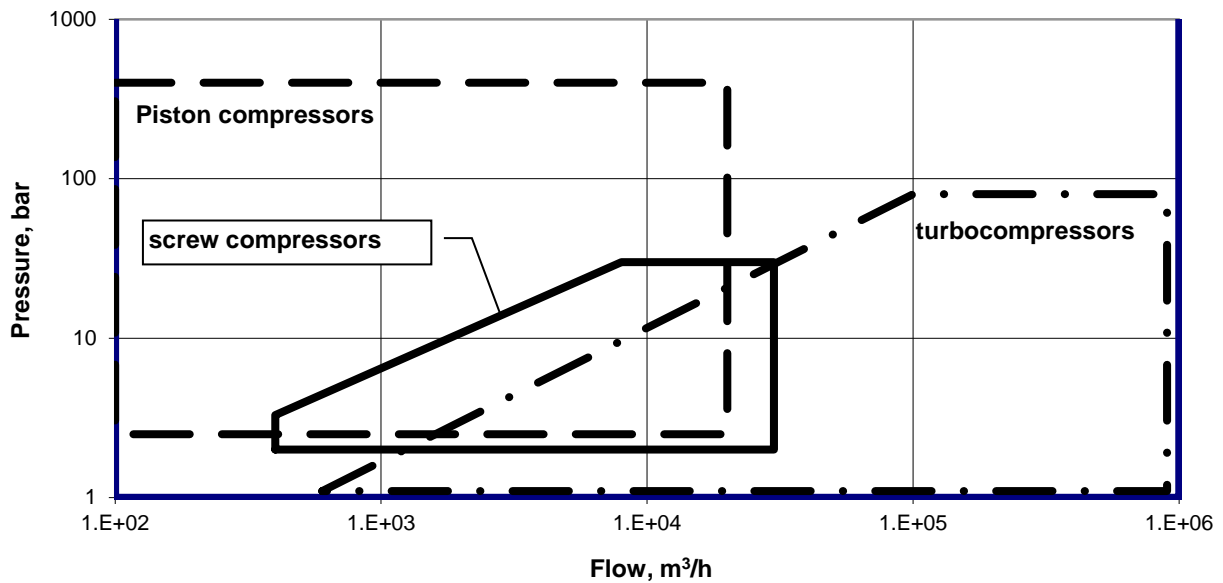

**Fig. 18:** Application ranges for different compressors

As a result, only piston and screw compressors are used for helium compression. Figure 18 shows the limitations for three types of compressors.

*The screw compressor*

"The twin-screw type compressor is a positive displacement type device that operates by pushing the working fluid through a pair of meshing close-tolerance screws similar to a set of worm gears ... Each rotor is radially symmetrical, but laterally asymmetric [see Fig. 19] ... The working area is the inter-lobe volume between the male and female rotors. It's larger at the intake end, and decreases along the length of the rotors until the exhaust port. This change in volume is the compression. The intake charge is drawn in at the end of the rotors in the large clearance between the male and female lobes. At the intake end the male lobe is much smaller than its female counterpart, but the relative sizes reverse proportions along the lengths of both rotors (the male becomes larger and the female smaller) until (tangential to the discharge port) the clearance space between each pair of lobes is much smaller. This reduction in volume causes compression of the charge before being presented to the intake manifold."[4]

All helium screw compressors are oil-lubricated machines. Here, compressor oil is injected into the compression cavities: it bridges the space between the rotors, both providing a hydraulic seal and transferring mechanical energy between the driving and the driven rotor. However, the most important function of the compressor oil is to provide an efficient cooling sink for the hot gas. The oil is separated from the discharge stream, then cooled, filtered and recycled.

The screw compressor is a low-cost device, because this kind of compressor is a bulk commodity: they are used to supply compressed air for general industrial applications or in conventional refrigeration. Compressed air screw compressors are suitable for helium compression in particular, because of the similar pressures (1 bar at the inlet) and pressure ratio (usually 8–10).

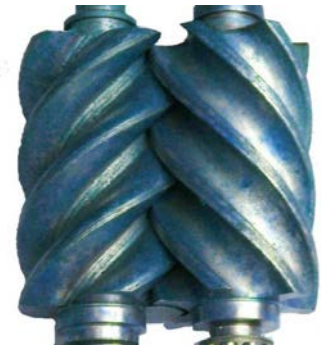

**Fig. 19:** The screws of a screw compressor

*Reciprocation (piston) compressors*

Another kind of compressor used in helium refrigeration is the reciprocating compressor. It is used for small-scale liquefaction (< 30 l/h), as the main helium compressor or for the compression of helium to high pressure (>30 bar), because screw compressors are not available for this pressure range.

### 4.2.2    *The expansion turbine (or expander or turbine)*

It is said that the heart of a helium liquefier is the expansion turbine. This means that an essential part of the know-how (in helium refrigeration) is about how to build an efficient, reliable, and service-friendly expansion turbine.

During the expansion from high pressure to low pressure, the gas produces mechanical power; the enthalpy of the gas (and the temperature) at the outlet is therefore lower than at the inlet. Theoretically, the mechanical power produced can be integrated into the refrigeration process. However, in reality this process design option is very uncommon – engineers prefer to dissipate this energy (transfer it by friction to the heat and reject it into the ambient) rather than using it. This is because in this way the refrigeration process becomes simple and easy to control; additionally, it is a slightly less expensive option. All these advantages justify the energy penalty of approximately 0.3–0.5%.

---

[4] From Wikipedia, 'Rotary screw compressor': see http://en.wikipedia.org/wiki/Rotary_screw_compressor

Figure 20 shows three different turbines in terms of bearing supports. The common features are as follows.

- The impeller is connected to the shaft, which is supported by two bearings at least.
- The operational temperature of the impeller is relatively low and the bearings work at temperatures close to ambient or higher; therefore, the distance between these two parts has to be as long as possible to avoid heat leakage due to conductivity through a solid.
- The two areas (the bearing and the impeller) are separated spatially to avoid cross-contamination between the main process gas stream and the portion of helium inside the bearing chamber.
- The impeller of a turbine with gas bearings is usually located below the bearing chamber; in contrast, the bearing part of a turbine with oil bearings is located below the impeller. This is necessary to prevent contamination of the impeller by the oil.

*Oil bearing (left)*

The oil bearing is the option developed in the early 1960s: the oil film guarantees the load-bearing capacity for both axial and radial loads. The oil is pumped to a high pressure by a separate oil pump and injected into the bearing gap. Later, it flows down to the bottom of the bearing chamber and further to a small vessel placed below. This vessel is required for separation of the helium gas and oil. It works simultaneously as a feed vessel for the oil pump. The helium gas in the bearing space appears from the main process helium stream, through the gap between the impeller space and the bearing chamber. This gas is then provided to the suction line of the main helium compressor. It is a kind a parasitic bypass stream, which affects the efficiency of the whole cycle.

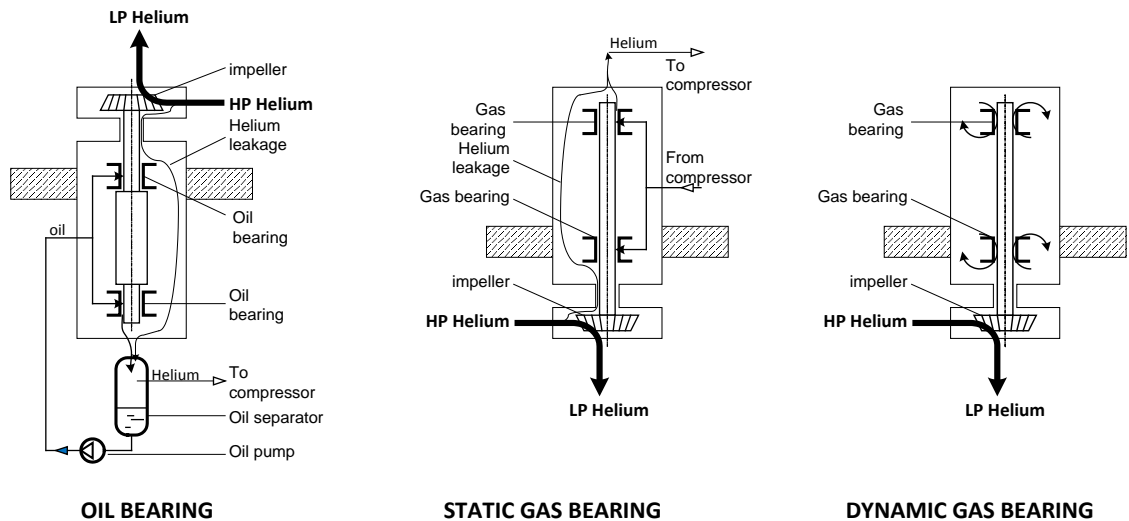

**Fig. 20:** Typical bearing systems

*Gas bearing*

Later, another solution was found by smart design engineers, based on the use of a gas film instead of the oil film in the bearing.

First, the so-called 'static gas bearing' was introduced to the market: here, the compressed helium gas from an external source (usually the main helium compressor) is injected into the tiny gap between the shaft and the support. The problem of potential contamination of helium by oil was elegantly solved in this way.

Subsequently, the so-called 'dynamic bearing' was invented. Here, the external high-pressure gas source and corresponding bypass stream are completely eliminated, on the basis of an appropriate design of the bearing, which allows the required pressure inside the bearing chamber to build up internally. The efficiency of gas expansion turbines with dynamic bearings is therefore a little higher in comparison to turbines with oil bearings and static gas bearings.

Figure 21 shows a modern turbine expander with dynamic gas bearings.

- The high-pressure feed gas enters the unit bottom-up (via two nozzles, on the left and the right), it expands in the impeller and exits the unit axially downwards (via the nozzle in the middle).
- The impeller of a small turbocompressor is mounted on the turbine shaft (above the yellow part in Fig. 21); therefore the turbine impeller drives this turbocompressor, which is part of a secondary closed cycle (the so-called 'braking cycle'): the gas compressed here becomes warm (as result of the compression), the heat is transferred to the cooling water in a compact heat exchanger (the water cooler), and the cooled gas is then throttled in a valve and flows again to the compressor inlet. In this way, the mechanical power produced by the turbine is dissipated into the ambient (the cooling water).

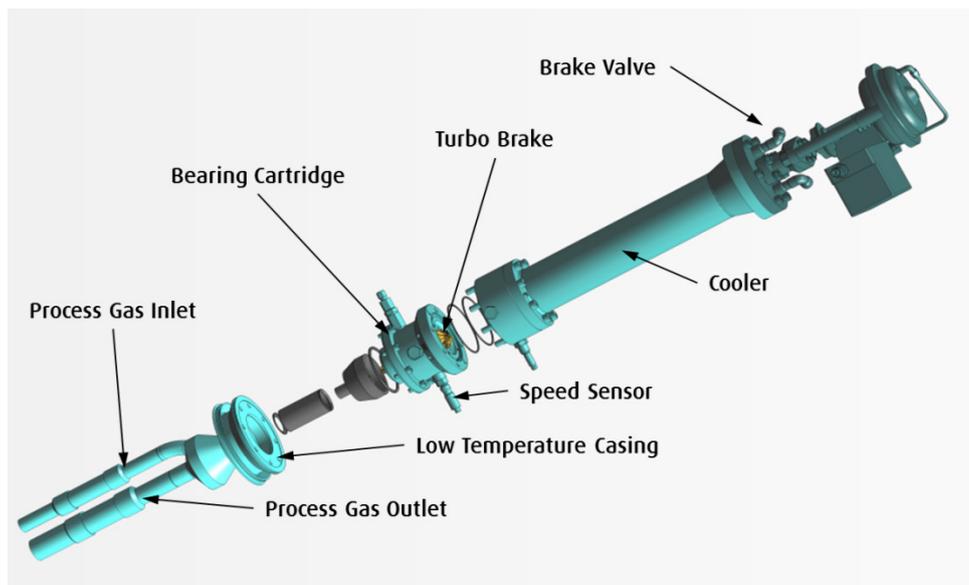

**Fig. 21:** An expansion turbine with dynamic bearings (Linde)

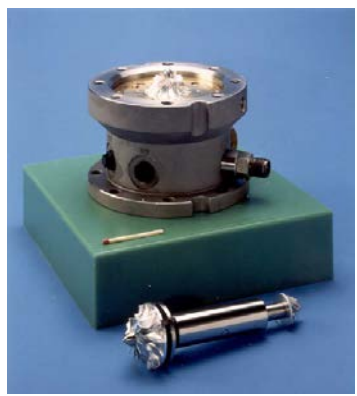

**Fig. 22:** A typical turbine cartridge

### 4.2.3 Heat exchangers

The heat exchangers in the cryogenic section are almost exclusively of the aluminium plate–fin type.

Figures 23 and 24 "show the structure of a plate fin heat exchanger module: the process streams are led through passages. Up to 200 of these passages are stapled one on top of the other. The large number of passages makes it possible to bring several streams into thermal contact within one unit. The outlet frame is formed by 10–25 mm side bars, which are only interrupted for passages inlets and outlets. A fluid enters the passage via nozzles and headers. Beginning from here the flow is distributed with special fins over the entire cross-section of the passage and passed to the main section with heat transfer fins. The arrangement of the passages as well as fin types can be selected by process design engineer according to the process requirements" [4].

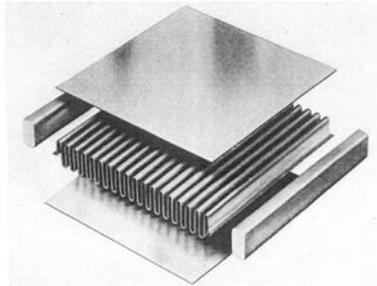

**Fig. 23:** Inside the plate–fin heat exchanger

The advantages of plate–fin heat exchangers are as follows:

– high flexibility concerning the number of process streams: several of them can be passed through the single block, which allows more sophisticated process design;
– a high specific heat exchange surface, which helps to realize a very efficient process because of the small temperature difference and pressure losses;
– low pressure losses;
– low specific costs.

**Table 4:** Typical parameters of aluminium plate–fin heat exchangers

| Parameter | Value(s) |
| --- | --- |
| Sizes | Up to 1.8 m × 1.5 m × 8.0 m |
| Specific surface | 500–1800 m²/m³ |
| Fin | 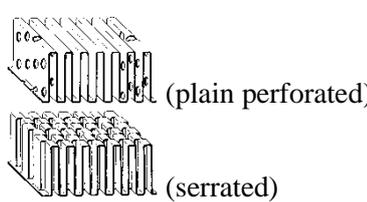 (plain perforated) <br> (serrated) <br> Fin height: 4–10 mm <br> Fin thickness: 0.1–0.6 mm |
| Temperature | –269°C to +65°C |
| Pressure | < 115 bar |
| Materials | ASTM 3003 (DIN AlMnCu), ASTM 5083, 6061 (DIN AlMg4.5Mn) |

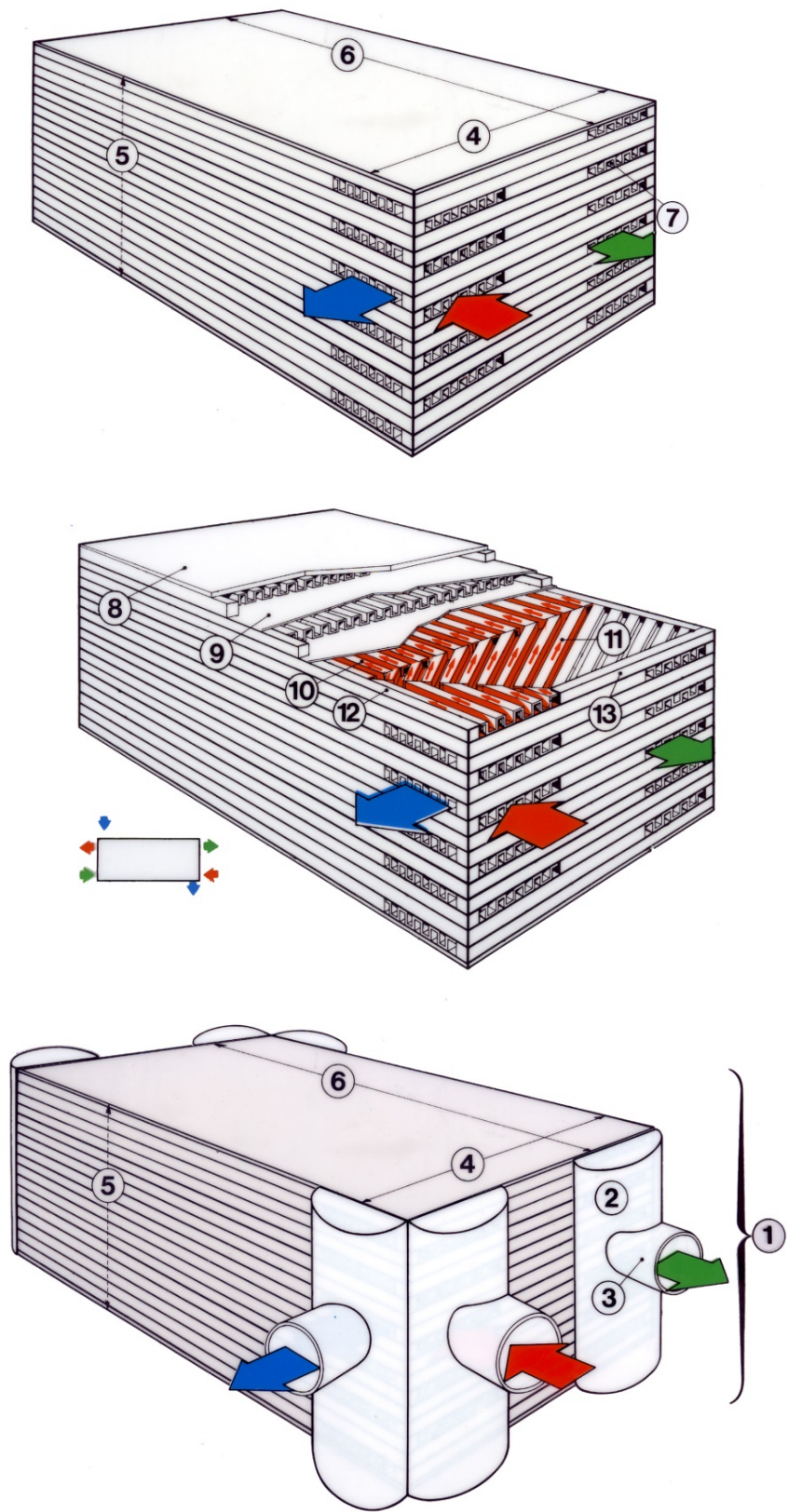

**Fig. 24:** An aluminium plate–fin heat exchanger: 1, core; 2, header; 3, nozzle; 4, width; 5, height; 6, length; 7, passage outlet; 8, cover sheet; 9, parting sheet; 10, heat transfer fins; 11, distribution fins; 12, side bar; 13, front bar.

Figure 25 shows a typical laboratory helium liquefier for universities and academic insitutions. It consists not only of the liquefier itself, but of all of the necessary infrastructural hardware.

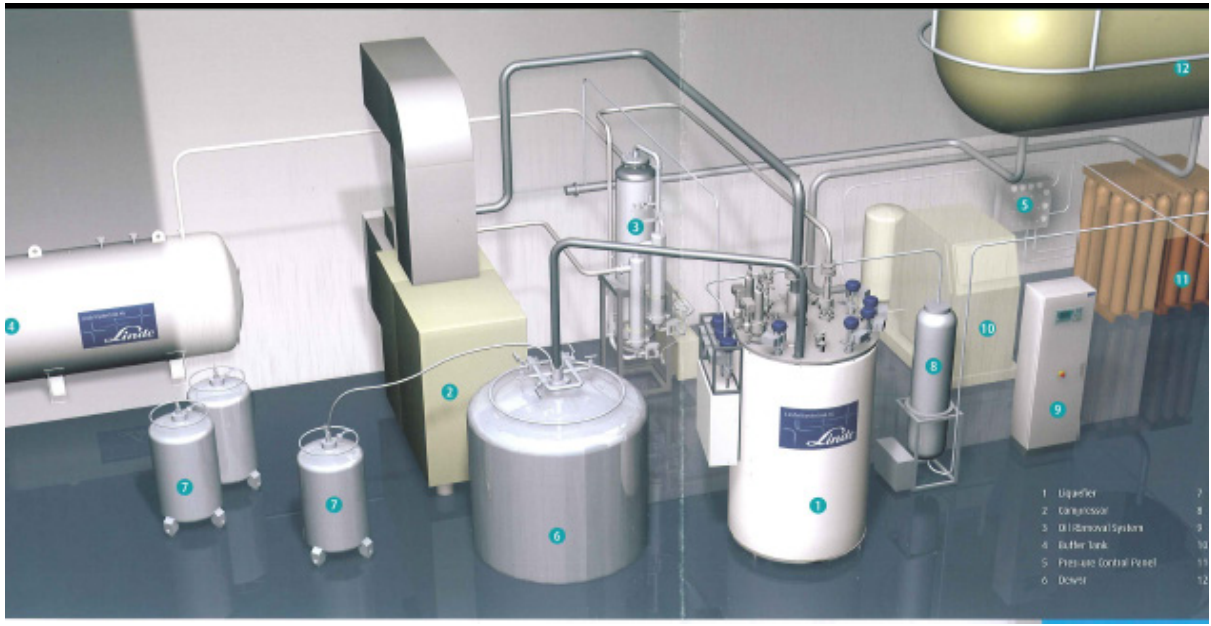

**Fig. 25:** A laboratory helium system: 1, liquefier; 2, main compressor; 3, oil separator; 5, pressure control panel; 6, main Dewar for LHe; 7, small Dewars; 8, dryer/purifier; 9, instruments; 10, recovery compressor; 11, high-pressure gas storage; 12, low-pressure gas storage.

Usually, the liquid helium is distributed by relatively small, compact transport Dewar vessels. This method is used for cooling purposes in numerous experiment set-ups, and consequently liquid helium evaporates during usage. The gas (contaminated by air and humidity) is usually collected in low-pressure helium balloons (which look very similar to conventional air balloons – attempts are made to use rubberized materials to prevent diffusion of the helium through the balloon wall). This gas is compressed to 150–200 bar with the aid of a small high-pressure compressor (recovery compressor) and transported back to the liquefier station. The amount of helium recovered can reach relatively high values, up to 90%, depending on the experience and carefulness of the helium users.

The small transport Dewars are also passed back to the liquefier: the best of them will have a small portion of liquid helium inside, to guarantee low temperature and minimize cooling downtime and energy.

The returned helium gas is dried, purified, liquefied, and transferred into the main storage vessel, which is a part of the liquefier system. The small Dewars are charged from this vessel on demand.

## 4.3 Trends

Karl Loehlein [5] has found the following trends concerning helium refrigeration.

- The requirements concerning the capacity of a single helium refrigeration unit are growing, particularly with regard to giant projects in high-energy physics, and specially fusion (ITER).
- The demand for refrigeration at a cooling temperature of 1.8 K is also growing.
- Efficiency is becoming increasingly important at higher cooling capacities.

- However, the requirements with regard to the efficiency of small liquefaction systems are also becoming higher.
- Progress in the development of control and instrumentation is leading to a higher degree of automation and to simplification of the operation of complex systems.